\documentclass[useAMS,usenatbib,fleqn]{mn2e}

\usepackage{times}
\usepackage{graphicx}
\usepackage{amssymb}
\usepackage{amsmath}
\usepackage{subfig}
\usepackage{url}
\usepackage{xspace}
\usepackage{gensymb}
\usepackage{enumerate}

\newcommand{\Msol}{M_{\odot}}
\newcommand{\Lya}{Ly$\alpha$\xspace}

\newcommand{\xm}{\langle x_i \rangle_\mathrm{m} }
\newcommand{\Mhalo}{M_{\mathrm{h}}}
\newcommand{\ud}{\mathrm{d} }
\newcommand\ion[2]{#1$\,${\scshape{#2}}}%
\newcommand\hi{\ion{H}{i}\xspace}
\newcommand\hii{\ion{H}{ii}\xspace}

\newcommand{\ctwo}{\textsc{C}$^2$\textsc{-ray}\xspace}
\newcommand{\cthree}{\textsc{CubeP}$^3$\textsc{M}\xspace}
\newcommand{\scic}{\sigma_{\mathrm{cic}}}

\title[Large \Lya emitter surveys]{Studying reionization with the next generation of \Lya emitter surveys}
\author[H. Jensen et al.]{H. Jensen$^{1}$\thanks{hannes.jensen@astro.su.se}, M. Hayes$^{1}$, I. T. Iliev$^{2}$, P. Laursen$^{3}$, G. Mellema$^{1}$, E. Zackrisson$^{1}$ \\
$^{1}$Dept.\ of Astronomy and Oskar Klein Centre, Stockholm University, AlbaNova, SE-10691 Stockholm, Sweden\\
$^{2}$Astronomy Centre, Department of Physics \& Astronomy, Pevensey II Building, University of Sussex, Falmer, Brighton BN1 9QH, UK\\
$^{3}$Dark Cosmology Centre, Niels Bohr Institute, University of Copenhagen, Juliane Maries Vej 30, 2100 Copenhagen, Denmark\\
}%\\
\begin{document}

\date{\today}

\pagerange{\pageref{firstpage}--\pageref{lastpage}} \pubyear{2013}

\maketitle

\label{firstpage}

\begin{abstract}

We study the prospects for constraining the ionized fraction of the intergalactic medium (IGM) at $z>6$ with the next generation of large \Lya emitter surveys. We make predictions for the upcoming Subaru Hyper Suprime-Cam (HSC) \Lya survey and a hypothetical spectroscopic survey performed with the James Webb Space Telescope (JWST). Considering various scenarios where the observed evolution of the \Lya luminosity function of \Lya emitters at $z>6$ is explained partly by an increasingly neutral IGM and partly by intrinsic galaxy evolution, we show how clustering measurements will be able to distinguish between these scenarios. We find that the HSC survey should be able to detect the additional clustering induced by a neutral IGM if the global IGM neutral fraction is greater than $\sim$20 per cent at $z=6.5$. If measurements of the \Lya equivalent widths (EWs) are also available, neutral fractions as small as 10 per cent may be detectable by looking for correlation between the EW and the local number density of objects. In this case, if it should turn out that the IGM is significantly neutral at $z=6.5$ and the intrinsic EW distribution is relatively narrow, the observed EWs can also be used to construct a map of the locations and approximate sizes of the largest ionized regions. For the JWST survey, the results appear a bit less optimistic. Since such surveys probe a large range of redshifts, the effects of the IGM will be mixed up with any intrinsic galaxy evolution that is present, making it difficult to disentangle the effects. However, we show that a survey with the JWST will have a possibility of observing a large group of galaxies at $z\sim7$, which would be a strong indication of a partially neutral IGM.
\end{abstract}

\begin{keywords}
	cosmology: dark ages, reionization, first stars---methods: numerical
\end{keywords}
%%%%%%%%%%%%%%%%%%%%%%%%%%%%%%%%%%%%%%%%%%%%%%

\section{Introduction}
\label{:sec:introduction}
The Epoch of Reionization (EoR) is the time period in the history of the Universe when the first sources of light formed and ionized the intergalactic medium (IGM). Observations of quasar spectra indicate that this process must have been largely completed around $z\approx6$ \citep{fan2006}, but the exact details beyond this redshift---both in terms of the timing of reionization and the sources responsible---are still not well known. Some additional observational constraints come from the polarisation of the cosmic microwave background \citep{komatsu2011,larson2011} and temperature measurements \citep{theuns2002,raskutti2012}. However, these observations are still consistent with a great range of reionization scenarios \citep{mitra2012}.

During the past decade, \Lya emitters have attracted considerable attention as potential probes of the latest stages of the EoR (e.g.\ \citealt{hu2002,malhotrarhoads2004,kashikawa2006,shimasaku2006,ota2008,stark2010,ouchi2010,hu2010,kashikawa2011,pentericci2011,caruana2013,treu2013,pentericci2014,tilvi2014}). Since \Lya photons are resonantly scattered by neutral hydrogen, the \Lya luminosity from a galaxy is suppressed if the galaxy is surrounded by a neutral IGM. Therefore, we would expect the statistics and number counts of \Lya emitters to change at redshifts where reionization is not yet completed. In principle, it should be possible to use this change to measure how neutral the IGM is at a certain redshift.

While some earlier studies saw no evidence of evolution in the \Lya luminosity function of \Lya emitters at redshifts $\gtrsim 6$ \citep{malhotrarhoads2004,tilvi2010}, in the past few years several observational studies have established that there is in fact a rapid evolution in the population of \Lya emitters. \cite{ouchi2010} and \cite{kashikawa2011} showed that the luminosity function drops in amplitude between $z=5.7$ and $z=6.5$, after being roughly constant in the interval $3 < z < 5.7$. This drop appears to continue at even higher redshifts \citep{konno2014}. In a similar vein, \cite{stark2010}, \cite{pentericci2011,pentericci2014}, \cite{treu2013}, \cite{caruana2013} and \cite{faisst2014} all find that the fraction of drop-out selected galaxies that show \Lya emission decreases rapidly between $6 < z < 7$. It is tempting to view these observations as evidence of an increase in the IGM neutral fraction at $z > 6$.

However, simulations that try to quantify how high the IGM neutral fraction needs to be to account for this evolution at $z>6$ typically require very late and rapid reionization scenarios, with neutral fractions as high as 40 per cent at $z=6.5$ \citep{dijkstra2011,pentericci2011,foreroromero2012,jensen2013}. Such high neutral fractions are hard to justify physically, and seem incompatible with measurements of the IGM temperature \citep{raskutti2012} and \Lya line profile measurements \citep{ouchi2010}. Several ways of easing this tension between simulations and observations have been proposed, including intrinsic evolution in the \Lya emitters \citep{dayal2011,hutter2014,dijkstra2014}, sample variance in the observations \citep{taylor2013} and the presence small-scale high-density structures \citep{bolton2013}.

While measurements of the luminosity function and the \Lya emitter fraction cannot distinguish between intrinsic evolution and IGM evolution, the additional clustering of \Lya emitters due to a partially ionized IGM could be used as a much more robust probe of the ionization state of the IGM \citep{mcquinn2007,ouchi2010}. As seen in \cite{jensen2013}, this method requires sample sizes of several thousand objects to put meaningful constraints on reionization. Such large samples are not available today, but upcoming observations with the Hyper-Suprime Cam (HSC) on the Subaru telescope, will observe some 5500 narrow-band selected \Lya emitters over a total of 30 deg$^2$ at $z=6.5$ (M. Ouchi 2013, private communication). Furthermore, the James Webb Space Telescope (JWST) will perform deep spectroscopic surveys of \Lya emitters at high redshifts. However, to date, few systematic investigations have been performed regarding what new information about reionization that will be possible to extract from these observations, besides better measurements of quantities like the \Lya luminosity function and \Lya emitter fraction.

This paper is a first step towards providing such study. We investigate the extent to which the next generation of large-field \Lya emitter surveys---such as the HSC survey---and deep spectroscopic surveys---for example with the JWST---may be able to distinguish between intrinsic galaxy evolution and a neutral IGM, focusing mainly on clustering measurements, which are beyond the reach of today's surveys. Furthermore, we explore the possibilities of using such large \Lya samples to perform ``bubble-mapping'', i.e. determining the positions and shapes of the largest \hii regions.

The paper is structured as follows: in section \ref{sec:methods} we describe our simulations and physical assumptions. We also discuss how the clustering of \Lya emitters is affected by a partially neutral IGM and how this clustering can be quantified. In section \ref{sec:results}, we present our predictions, first for the HSC survey and then for the JWST survey. Here, we also comment briefly on the recent observational results of \cite{pentericci2014} and \cite{tilvi2014}. Finally, in section \ref{sec:summary}, we summarize and discuss our results.

For the simulations, we have assumed a flat $\Lambda$CDM model with $(\Omega_{\mathrm{m}}, \Omega_{\mathrm{b}}, h, n, \sigma_8) = (0.27, 0.044, 0.7, 0.96, 0.8)$, consistent with the 9 year WMAP results \citep{hinshaw2012}.

%%%%%%%%%%%%%%%%%%%%%%%%%%%%%%%%%%%%%%%%%%%%%%

\section{Methods}
\label{sec:methods}

Our goal is to produce mock observational samples that incorporate different combinations of intrinsic galaxy evolution and evolution of the IGM ionized fraction. To do this, we start with a large cosmological simulation of the IGM. We then assign intrinsic \Lya emission properties to the dark matter haloes in the simulation and calculate the \Lya emission through the IGM at various global ionized fractions. The methods used here are essentially identical to those described in detail in \cite{jensen2013}, although we use a much larger simulation volume here.

\subsection{Simulations of the IGM}
To simulate the reionization of the IGM we start with a (607~cMpc)$^3$ $N$-body simulation produced with the \cthree code \citep{harnoisderaps2012}. Based on \textsc{PMFAST} \citep{merz2005}, \cthree works by calculating gravitational forces on a particle-to-particle basis at short distances, and on a grid at long distances. For this simulation, we used 5488$^3$ particles on a 10976$^3$ grid, which was eventually down-sampled to 504$^3$ grid points. The particle mass was $5 \times 10^7 \Msol$.

After identifying haloes with a spherical-overdensity halo finder, we simulate the reionization of the IGM using the \ctwo code \citep{mellemac2ray}. \ctwo is a photon-conserving ray-tracing code that uses short-ray characteristics to solve the equation for the evolution of the ionization fraction of hydrogen by iterating over all sources and grid points until convergence.

In this simulation, we assigned all haloes an ionizing flux $\dot{N}_{\gamma}$ proportional to the halo mass $\Mhalo$:
\begin{equation}
	\dot{N}_{\gamma} = g_{\gamma} \frac{M_{\mathrm{h}} \Omega_\mathrm{b}}{(10\;\mathrm{Myr}) \Omega_\mathrm{m} m_{\mathrm{p}}},
	\label{eq:ionizing_flux}
\end{equation}
where $m_{\mathrm{p}}$ is the proton mass and $g_{\gamma}$ is a source efficiency coefficient that effectively incorporates the star formation efficiency, the initial mass function and the escape fraction. The haloes identified with the halo finder had masses down to $10^9 \; \Msol$; these were assigned a source efficiency of $g_{\gamma} = 1.7$. In addition to this, we also added smaller sources with $\Mhalo$ down to $10^8\;\Msol$ using a subgrid recipe (Ahn et al., in preparation). These were assumed to have a $g_{\gamma} = 7.1$, to account for a lower metallicity and a more top-heavy initial mass function. On the other hand, since small haloes lack the gravitational potential to keep accreting matter in a highly ionized environment (e.g.\ \citealt{iliev2007}), these low-mass sources were turned off when the local ionization fraction exceeded 10 per cent. These source efficiencies were chosen to give a reionization history that is consistent with observational results for the electron-scattering optical depth and quasar spectra. For more information about these codes and this particular simulation run and the assumptions behind the simulation parameters, see \cite{iliev2013}.

\subsection{Simulations of \Lya emitters}
\label{sec:simulations}
The next step is to simulate the observed \Lya emission from the haloes from the previous step. We use the same two-step model as in \cite{jensen2013}, where we model the \Lya line shape emerging from the circum-galactic medium, $J_{\mathrm{em}}(\lambda)$, using a recipe based on high-resolution simulations, and then trace sight-lines through the IGM, calculating the optical depth due to scatterings out of the line-of-sight. The division between the circum-galactic and extragalactic parts was taken to be at $1.5r_{\mathrm{vir}}$, where $r_{\mathrm{vir}}$ is the virial radius of the halo. This value is motivated by \cite{laursen2011}, whose simulations show that at $z=6.5$, 80 per cent of \Lya photons have experienced their last scattering into the line-of-sight.

The line shape recipe we use is the Gaussian-minus-Gaussian (GmG) model introduced in \cite{jensen2013}. This recipe is a fit to high-resolution hydrodynamical+\Lya radiative transfer simulations. While there is no physical justification for this recipe, it has turned out to match the high-resolution simulations well. The GmG model gives double-peaked line shapes that are symmetric on the blue and red sides of the line centre, and widen as the halo mass increases. Such a line shape gives a slightly weaker dependence on IGM neutral fraction for the IGM transmission of \Lya than models where a big part of the emission comes from the line centre (e.g.\ \citealt{dijkstra2011,jensen2013}). This will tend to make our predictions here somewhat on the conservative side. In reality, far from all observed \Lya lines show a double-peaked structure. However, at the redshifts considered here, the blue peak will almost always be absorbed by the IGM, and will only affect the normalization of the intrinsic \Lya luminosity (see below). For a more in-depth discussion about other line-shape models, both analytical and simulated, see \cite{jensen2013}.

For the IGM part of the \Lya radiative transfer, we use the same modified version of \textsc{IGMtransfer} \citep{laursen2011} as described in \cite{jensen2013}. The optical depth for \Lya photons is summed over all cells along the sight-line:
\begin{equation}
	\tau(\lambda) = \sum_i^{\mathrm{cells}} n_{\mathrm{HI,i}} \sigma_{\mathrm{Ly}\alpha}(\lambda + \lambda v_{||,i}/c) \Delta r,
	\label{eq:optical_depth}
\end{equation}
where $\Delta r$ is the length of each step, $n_{\mathrm{HI,i}}$ is the local neutral hydrogen density, $\sigma_{\mathrm{Ly}\alpha}$ is the neutral hydrogen cross section to \Lya scattering and $v_{||,i}$ is the gas velocity component of the cell along the line of sight, including the Hubble flow. Using this method, we calculate the transmission of \Lya through the IGM for sightlines along the three coordinate axes for all haloes with $\Mhalo > 10^{10} \; \Msol$. By multiplying the line profile with $\exp(-\tau_{\lambda})$ we then get the observed line profile. The fraction $T_{\alpha}$ of \Lya photons that are transmitted through the IGM is then:
\begin{equation}
	T_{\alpha} = \frac{\int J_{\mathrm{em}}(\lambda)\exp(-\tau_{\lambda}) \ud \lambda}{\int J_{\mathrm{em}}(\lambda) \ud \lambda}.
	\label{eq:transmitted_fraction}
\end{equation}

A weakness of this method is that the division between the circum-galactic and extragalactic part (which we put at $1.5$ virial radii; see discussion above) in reality depends on the amount of neutral hydrogen. While our value appears reasonable for a fairly ionized medium, it is probably less realistic for a more neutral case. Ideally, one would carry out the full radiative transfer of \Lya for every galaxy out to a distance where no more photons scatter into the line-of-sight. However, this would require a very high resolution and is computationally unfeasible. In practice, all studies of the effect of the IGM on \Lya emitters have to make some simplified assumptions. Here, we therefore use the same division and line model for all ionization fractions, but note that this may cause us to underestimate the effects of the IGM somewhat.

%A more proper approach would be to carry out the full radiative transfer of \Lya for every galaxy out to a distance where no more photons scatter into the line-of-sight. However, this would require a very high resolution and would be very computationally expensive. Here, we therefore use the same division and line model for all ionization fractions, but note that this may cause us to underestimate the effects of the IGM somewhat.

We also need a model for the intrinsic \Lya and UV continuum luminosities of the galaxies (here, we use the word \emph{intrinsic} to mean the luminosity of the light as it exits the galaxies at $1.5r_{\mathrm{vir}}$). Since we are interested only in the evolution of the observables due to the IGM neutral fraction, we make no attempt to construct a physically motivated model for the intrinsic properties. Instead, we simply calibrate the luminosities to observations at $z=5.7$.

In our model for the intrinsic properties, each dark matter halo with mass $\Mhalo$ from the $N$-body simulations is assigned a \Lya luminosity randomly drawn from a log-normal distribution with a standard deviation of $\sigma=0.4$ dex and a mean that is proportional to $\Mhalo$. The normalisation of the luminosities is simply a fit to observations and has no influence on our results, since we are only interested in the changes of various properties. For example, the normalisation of the intrinsic \Lya luminosity is degenerate with our line shape model. Had we assumed a line shape with no red peak, the intrinsic \Lya luminosities would have been a factor 2 lower. The change in luminosity with an increasing IGM neutral fraction, however, would be the same. 

The more important quantity here is $\sigma$, which gives us the amount of random scatter in the distributions of luminosities. More random scatter leads to a weaker correlation between galaxy mass and luminosity, which, as we shall see, makes methods that rely on the spatial clustering of \Lya emitters less viable for constraining reionization. The value of $\sigma$ was chosen based on the same set of high-resolution simulations that we used for the \Lya line shape model (see \cite{jensen2013} for more details). As we shall see in the next section, our \Lya luminosity model results in luminosity functions with rather shallow slopes compared to observations. Since higher values of $\sigma$ give shallower slopes, this indicates that if anything, we are over-estimating the random scatter in the \Lya luminosities. This would make our predictions for clustering somewhat conservative, since more randomness tends to make luminosity-selected galaxies appear less clustered.

\begin{figure}
	\centering
		\includegraphics[width=\columnwidth]{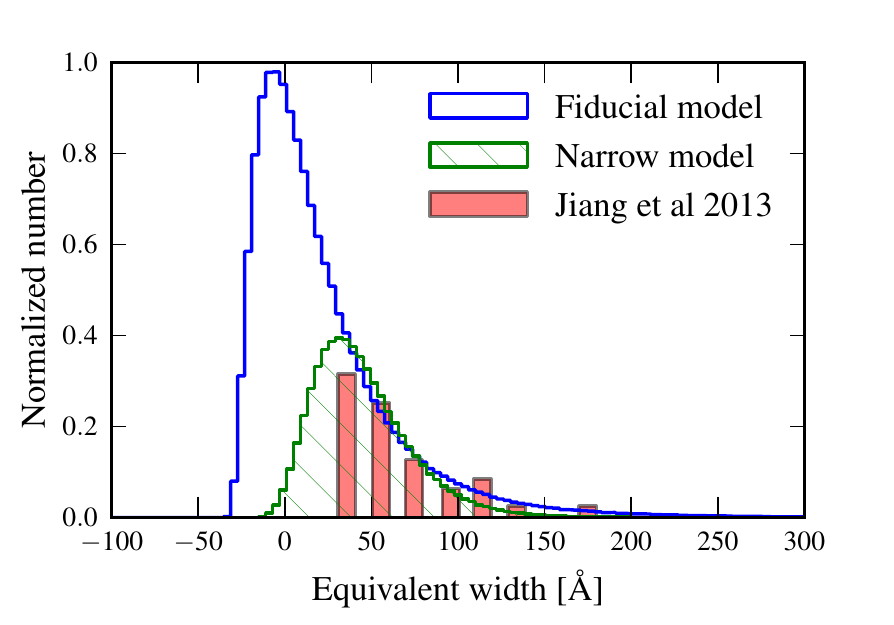}
		\caption{The probability density function for our two assumed EW distributions, compared with the observations by \protect\cite{jiang2013}. For the rest of the paper, results are shown for the fiducial model, unless stated otherwise. The distributions shown here are those which would be observed assuming 50 per cent transmission of \Lya through the IGM.}
	\label{fig:rew_distribution}
\end{figure}

For the \Lya EW---essentially the ratio between \Lya and UV continuum luminosity---we use a similar approach, assigning each dark matter halo an EW drawn from a probability distribution (for the remainder of the paper, we will use ``EW'' to mean restframe EW). The exact shape of the intrinsic EW distribution at $z=6.5$ is not well known, but observations from lower redshifts suggest that a log-normal distribution is reasonable (e.g.\ \citealt{reddy2009}). Our fiducial EW distribution is shown as the blue histrogram in Fig.\ \ref{fig:rew_distribution}. In this model, 65 per cent of all galaxies have \Lya EWs lower than 25 \AA, which is consistent with the EW distribution of Lyman-break selected galaxies observed by \cite{jiang2013}. Later on in the paper, we also consider a more narrow EW model, shown as a green histogram in Fig.\ \ref{fig:rew_distribution}. In the narrow model, approximately 25 per cent of all galaxies have a \Lya EW lower than 25 \AA. This is only marginally consistent with the least luminous galaxies of \cite{stark2010}, and thus represents a bit of an extreme case. For the rest of the paper, if nothing else is stated, results are shown for the fiducial EW model. For both models, we assume that the EW distribution is valid at $z=5.7$ for 50 per cent IGM transmission, and that any change in the EW at higher $z$ is due to the IGM only. Further, we assign the EWs completely randomly to our galaxies, assuming that there is no correlation between \Lya luminosity and EW. This seems to be supported by observations, at least for objects with detectable EWs \citep{zheng2013,hayes2014}.

\subsection{Physical scenarios}
\label{sec:scenarios}
A large part of our focus in this paper is on the change in the \Lya emitter population between $z=5.7$ and $z=6.5$, since these are the highest redshifts where the HSC survey will obtain substantial samples. Previous observations have indicated that the \Lya luminosity function of \Lya emitters changes between these two redshifts \citep{ouchi2010,kashikawa2011}, but it is not clear whether this is due to an increasingly neutral IGM, a change in the intrinsic properties of \Lya emitters or a combination of the two. 

\begin{figure}
	\centering
		\includegraphics[width=\columnwidth]{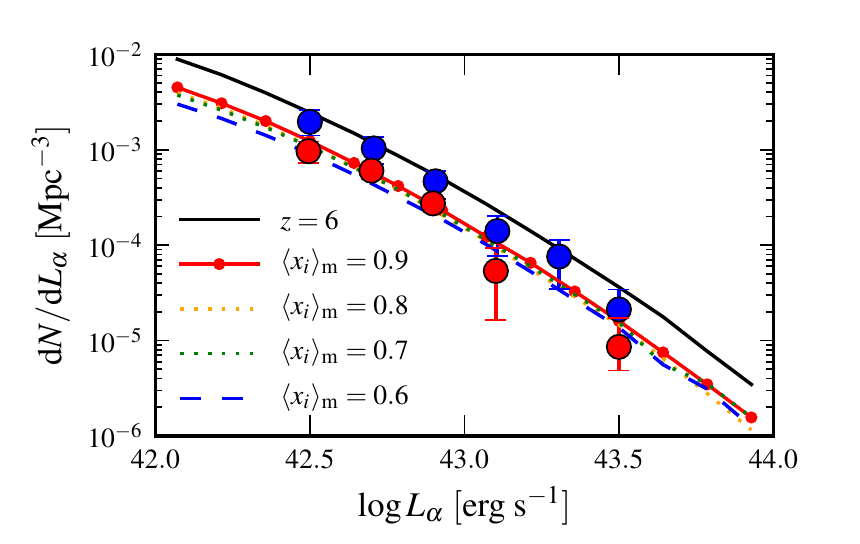}
        \caption{\Lya luminosity functions for the physical scenarios considered in the first part of this paper. The blue dots are observations by Ouchi et al at $z=5.7$. Red dots are observations by Ouchi et al 2010 at $z=6.5$. All the different simulated scenarios give luminosity functions that are indistinguishable from each other.}
	\label{fig:lfs_scenario}
\end{figure}

For this paper, we will make the assumption that the luminosity function at $z=6.5$ from \cite{ouchi2010} holds true. For the mock HSC observations in the first part of the paper, we consider a number of different scenarios where there are varying degrees of change in the IGM ionized fraction and the intrinsic properties of the \Lya emitters. There are several possible mechanisms that could change the intrinsic properties of \Lya emitters, including changes in dust content and star formation rate \citep{verhamme2006,laursen2009,guaita2011,bish2013,hayes2014}. Here, we do not specify the mechanism responsible. Instead, we simply model intrinsic evolution by dimming/brightening all objects in our mock samples by some constant amount. Effectively, this means that we dim/brighten the objects both in \Lya and UV continuum luminosity so that their intrinsic EWs are unchanged. One could also imagine intrinsic processes that would alter \Lya and UV luminosities in different ways. However, this distinction is unlikely to have any major effects on the statistical observables we discuss here, i.e.\ clustering of galaxies.

We label our scenarios ``$\xm=f$'', where $f$ is the global mass-averaged ionized fraction of the Universe at $z=6.5$. Our most extreme scenario is ``$\xm=0.6$'' (i.e.\ the IGM is only 60 per cent ionized at $z=6.5$). This is the ionized fraction we need in order to explain the change in the luminosity function without any intrinsic galaxy evolution between $z=5.7$ and $z=6.5$. Apart from this scenario, we also consider 70, 80 and 90 per cent ionization at $z=6.5$. In these scenarios, the rest of the luminosity evolution is due to intrinsic evolution (which we model by simply dimming all haloes by a constant factor). Later on, in section \ref{sec:jwst}, we will take a slightly different approach, and instead compare a scenario with simulated IGM evolution to two different models for intrinsic evolution.

Since we have only one reionization simulation, we have to construct these scenarios from simulation volumes that are nominally at different redshifts. For example, for the $\xm = 0.6$ scenario, we use the \ctwo output from $z=7.5$ and increase all the halo masses so that the intrinsic luminosity function matches the one at $z=5.7$. This is of course not entirely self-consistent, but since many studies have shown that the reionization topology is much more dependent on global ionized fraction than on redshift---at least if the relative contribution of high-mass and low-mass sources is similar---this method should still be reasonable \citep{mcquinn2007,mcquinn2007b,friedrich2011,iliev2012,taylor2013}. To verify that the different redshifts do not have an appreciable impact on the intrinsic properties of our galaxies, we have calculated both the \Lya luminosity functions and the counts-in-cells variances (see Sec.\ \ref{sec:clustering}) for the samples with the effects of the IGM removed. We found the differences between the samples to be negligible.

\begin{figure}
	\centering
		\includegraphics[width=\columnwidth]{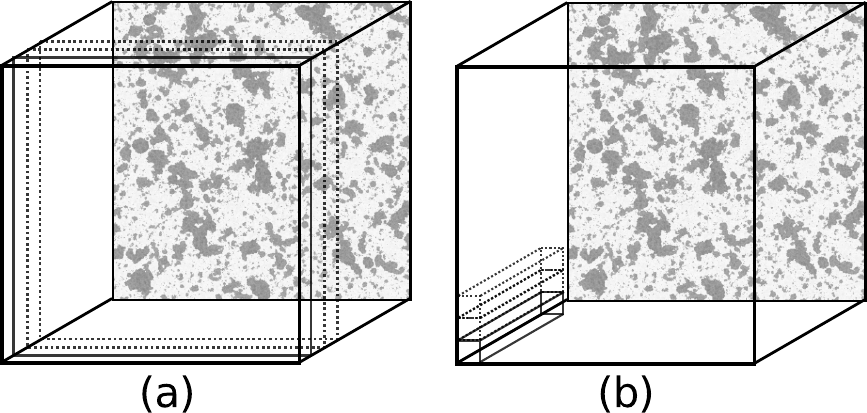}
	\caption{Schematic illustration of the sizes of our mock observations in relation to our full simulation volume. The left box (a) shows the HSC-like observations, where the width of the field is the full simulation volume, and the depth is $\Delta z\approx 0.1$. Here, we construct 10 samples along each coordinate axis. The right box (b) shows the JWST skewers, which have a field-of-view of 220 arcmin$^2$, and a depth of $\Delta z=1$. We construct a total of 256 such skewers from our full volume.}
	\label{fig:lae_boxes}
\end{figure}

For each of the scenarios, we construct a number of mock HSC samples by first selecting all objects with EW $>20$~\AA~, and then selecting the ones with the highest \Lya luminosities to give the same number density of \Lya emitters as the planned HSC survey (i.e.\ 5500 objects per 30 deg$^2$). This gives similar effective luminosity cutoffs as those expected for the HSC survey, but we chose to specify exact number densities instead of exact cutoffs to make clustering comparisons easier. Since the redshift depth of the HSC survey is approximately one tenth of our simulation volume, we are able to construct ten mock observations for each coordinate axis for each scenario. Each such mock observation will have roughly half the area of the full HSC survey, so for some statistical predictions we combine two separate mock observations. Fig.\ \ref{fig:lfs_scenario} shows the \Lya luminosity functions of all these mock samples along with the observations of \cite{ouchi2010} at $z=5.7$ and $z=6.5$. By design, they are indistinguishable when looking only at the luminosity functions. In Fig.\ \ref{fig:lae_boxes} we show a schematic illustration of how the mock samples compare to our full simulation volume and to the mock samples for the JWST survey that we consider in section \ref{sec:jwst}.

\begin{figure*}
	\centering
		\includegraphics[width=17cm]{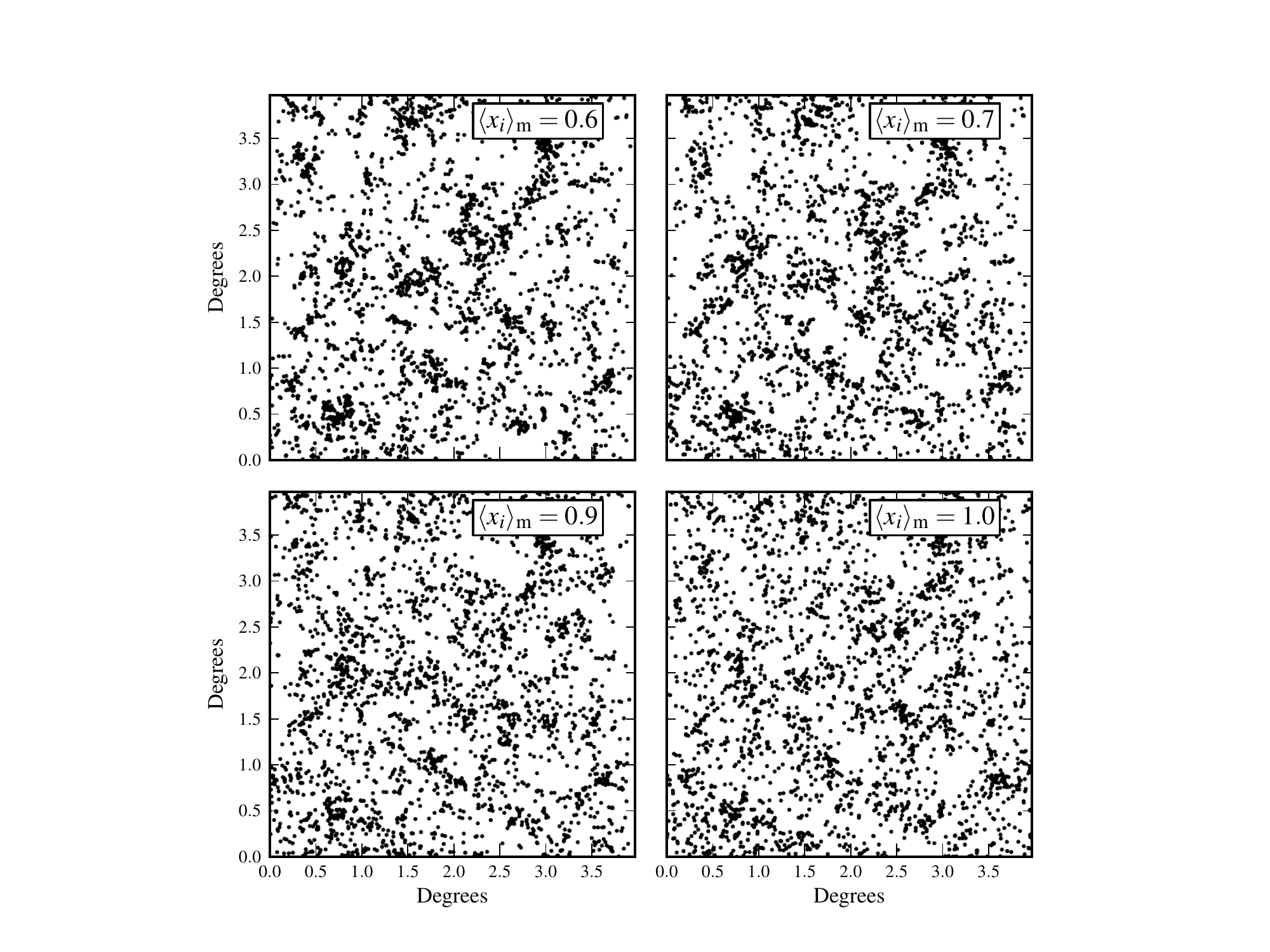}
	\caption{Observed \Lya emitters in a few sample mock HSC observations for some of our IGM evolution scenarios. The number density of objects in all panels is the same as expected for the HSC survey. However, the actual HSC field is roughly twice the area of our simulation. The additional clustering of objects in the less ionized scenarios is clearly visible.}
	\label{fig:observed_samples}
\end{figure*}

\subsection{Clustering measurements}
\label{sec:clustering}
As can clearly be seen in Fig.\ \ref{fig:lfs_scenario}, observations of luminosity functions alone are not useful for telling intrinsic galaxy evolution apart from IGM evolution. However, by looking at the spatial clustering of \Lya emitters, it may be possible to separate the two effects. When doing this, it is important to distinguish the clustering that comes from the topology of the ionized bubbles (which is what we are interested in here) from other sources of clustering. 

In addition to the intrinsic clustering of galaxies due to density fluctuations in the Universe, a partially neutral IGM will affect the clustering of \Lya emitters in two ways \citep{mcquinn2007,mesinger2008,iliev2008}. First, if one considers two observed samples with the same luminosity cutoff but different IGM ionized fractions, the sample with the most neutral IGM will have a higher clustering of sources. This is mainly because only the most massive sources will be seen through the more neutral IGM, and these are also the ones that are found in the densest regions.

Second, if one considers two samples again with different IGM ionized fraction, but with the same number of observed sources (implying different luminosity cutoffs), then the clustering of sources will also be increased since \Lya emitters are easier to observe if they are located deep inside an ionized region. High-luminosity \Lya emitters that happen to lie at the edge of an ionized region may be obscured, while slightly lower-luminosity sources that are in the middle of an ionized region may be observable. However, since the ionized regions are not located at random, but are localized around the largest clusters of galaxies, this effect is fairly weak and requires large samples to observe.

The first of these two effects is simply due to the decrease in number density, and cannot be used to separate intrinsic galaxy evolution from IGM evolution. The second one, on the other hand, is a signature of a partially ionized IGM. Therefore, we consider only samples with a fixed number density. A visual illustration of the IGM-induced clustering for a fixed sample size is shown in Fig.\ \ref{fig:observed_samples}, where we show the detected \Lya emitters in four of our scenarios scenarios for the mock HSC survey. 

There are several ways to quantify the spatial clustering of galaxies. One of the most common methods is the two-point correlation function, $w(\theta)$, which is the excess probability of finding two galaxies at a separation $\theta$, compared to a uniform random distribution. The two-point correlation function contains a lot of information since it measures the clustering over a range of scales. However, comparing two different correlation functions is usually done by fitting a parameterized functional shape, which requires implicit assumptions about the statistics of the objects in the sample. Furthermore, the two-point correlation function is awkward to calculate for fields with complicated shapes.

Here, we consider instead a related statistic called \emph{counts-in-cells}. The counts-in-cells method works by simply dividing the observed field into a number of cells of equal size and counting the number of galaxies $n_i$ that fall into each cell $i$. Typically, one then calculates the variance of the counts:
\begin{equation}
	\scic^2 =\frac{1}{N} \sum_{i=0}^{N} (n_i - \langle n \rangle)^2
	\label{eq:cic}
\end{equation}
where $N$ is the total number of cells. The counts-in-cells variance gives us a single, easy to calculate, number that measures the clustering without any assumptions of the statistics of the objects---higher variance means more clustered objects. In fact, $\scic^2$ can be shown to be proportional to the integral of the two-point correlation function (see e.g.\ \citealt{adelberger1998} and \citealt{mesinger2008} for more details). The two major drawbacks of the method are that the results depend on the choice of the cell size, and that it contains no information about the clustering on different spatial scales (as opposed to the correlation function).

\subsection{Bubble mapping}
\label{bubblemapping}
The counts-in-cells method discussed above requires only that we know the positions of \Lya emitters. However, if we know also the EW of the galaxies in our sample, we can use this to learn more about the IGM. Since the UV luminosity is not affected by neutral hydrogen in the IGM\footnote{It is true that the part of the continuum on the blue side of the \Lya line will be subject to IGM absorption, but at redshifts $z\gtrsim6$, this absorption is essentially a complete Gunn-Peterson trough, and can be corrected for. The presence of a \Lya damping wing will introduce a slight uncertainty to the correction, but for a broad-band measurement, this uncertainty will be small.}, the EW---i.e.\ the ratio between the \Lya and the UV continuum luminosity---will be a measure of $T_{\alpha}$, the fraction of \Lya flux that is transmitted through the IGM.

For an individual galaxy, calculating $T_{\alpha}$ from the EW is fairly meaningless, since the random scatter is very large (cf.\ Fig.\ \ref{fig:rew_distribution}). However, with a sample containing thousands of galaxies, it becomes possible to look for trends in the observed EW in different regions on the sky. We would expect the EW to be higher on average in regions where the IGM is highly ionized. By studying the mean EW in different regions, it may be possible to reconstruct the ionization structure of the IGM and measure the locations and sizes of the largest ionized regions (or ``bubbles''). Such ``bubble mapping'' may give valuable insights into the topology of reionization. This approach is different from so-called \Lya intensity mapping \citep{silva2013,pullen2014} in that it only deals with galaxies, not the diffuse emission from the IGM itself.

%%%%%%%%%%%%%%%%%%%%%%%%%%%%%%%%%%%%%%%%%%%%%%

\section{Results}
\label{sec:results}
The HSC survey will attempt to measure both the \Lya and UV continuum luminosities of \Lya emitters at $z=5.7$ and $z=6.5$. In total, the survey is expected to detect approximately 5500 objects at $z=6.5$ using the narrow-band selection technique (M. Ouchi 2013, private communication). In this section, we give predictions for how accurately this survey will be able to constrain reionization. We begin by looking at the case where only \Lya luminosities are available. In section \ref{sec:lya_plus_uv} we consider the case where we also know the UV luminosities. In section \ref{sec:jwst} we investigate what can be done with a spectroscopic survey carried out using the JWST, where we would know the redshifts of objects in a field with a small area, but a large extent along the line-of-sight.

\subsection{HSC observations with \Lya luminosities only}
\label{sec:lya_only}
If the only available information about a galaxy is its \Lya luminosity, it is impossible to say anything about the \Lya transmission through the IGM for that particular galaxy. However, we can still say something about the ionization state of the IGM through clustering analysis, as discussed in section \ref{sec:clustering}.

\begin{figure}
	\centering
		\includegraphics[width=\columnwidth]{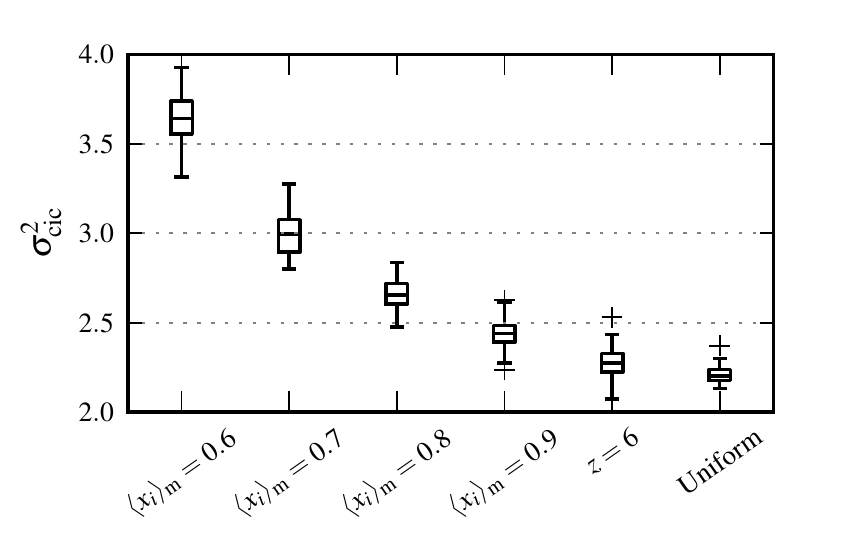}
		\caption{Counts-in-cells variances for our four scenarios and for the case of a uniform random distribution. The horizontal lines show the median of 30 different slices through the box. The boxes show the quartiles, the whiskers show the range of the data and the + signs show the outliers.}
	\label{fig:count_in_cells}
\end{figure}

In Fig.\ \ref{fig:count_in_cells}, we show the counts-in-cells variance $\scic^2$ for our four scenarios, using 50$\times$50 cells, i.e.\ a cell size of 4.8$\times$4.8 arcmin. This choice of cell size is somewhat arbitrary, but we have found that the results are not very sensitive to the exact cell size. Since our simulation area is half the size of the HSC field, we construct a mock measurement by combining two different slices of the volume. We do this for 30 different pairs of slices (we take 10 slices from the simulation volume along each coordinate axis, each with a depth of $\Delta z\approx 0.1$) to get an idea for the sample- and cosmic variance. The horizontal lines in Fig.\ \ref{fig:count_in_cells} show the medians of these 30 values of $\scic^2$. 

The actual values of $\scic^2$ are not important to us here. Instead, we are interested in the difference between the clustering at $z=6$---where we assume that the IGM is almost entirely ionized (e.g.\ \citealt{songaila2004})---and $z=6.5$. Comparing $\scic^2$ for samples with the same number of objects at these two redshifts, we may ask ourselves whether the change in clustering is consistent with a fully ionized IGM at $z=6.5$ or not. 

The variation in the measured $\scic^2$ comes both from Poissonian sample variance and from fluctuations in the underlying density field and the ionization fraction (cosmic variance). To better see the contribution of these effects, we performed the same calculations of $\scic^2$ as discussed above, but for uniformly distributed samples with the same number density as our mock observations (labelled ``Uniform'' in Fig.\ \ref{fig:count_in_cells}).

The variation in $\scic^2$ for the uniform samples is due only to sample variance. Comparing to the mock observations with a fully ionized IGM, we see that the variation is very similar in magnitude, indicating that the cosmic variance is negligible for such large samples, at least for the scenarios with higher ionization fraction\footnote{Note that the actual values of $\scic^2$ are lower since these data points are uniformly random, with no intrinsic galaxy clustering.}. For the $\xm = 0.6$ scenario, the variation in $\scic^2$ is roughly twice as large, due to the ionization fraction fluctuations.

Since the cosmic variance is negligible for a highly ionized IGM, any detection of a clustering signal that is higher than the $z=5.7$ clustering by more than the sample variance would be a strong indication of a significantly neutral IGM. Judging by Fig.\ \ref{fig:count_in_cells}, a sample with the size of the HSC survey can be used to robustly rule out a global ionization fraction of $\xm \lesssim 0.8$ at $z=6.5$. However, a neutral fraction as small as 10 per cent would be difficult to reliably detect with this method.

Translating from $\scic^2$ to $\xm$ of course assumes that our model for the \Lya emitters and the IGM is correct. There are a number of aspects of the model that can potentially affect the results. First, the model for the intrinsic \Lya line shape can make the observed luminosity more or less dependent on the surrounding IGM, as discussed in more detail e.g.\ by \cite{dijkstra2011} and \cite{jensen2013}. In general, the larger the part of the \Lya line that is offset from the line center, the less sensitive the transmission will be to the damping wing from the IGM, and the smaller the clustering due to neutral IGM will be. The line model we use here has no emission directly at the line center, and is thus relatively insensitive to the IGM. The intrinsic random scatter that we include in the \Lya luminosity will also affect the clustering: the more random the luminosities, the smaller the relative effect of the IGM on the object selection will be. The random scatter we include effectively includes both the randomness in the amount of \Lya that is produced in galaxies and the amount that escapes in different directions. These quantities are not well known, but looking at Fig.\ \ref{fig:lfs_scenario} it seems clear that including a larger random scatter than the one we use here would make the \Lya luminosity function too flat to be compatible with observations (since more random scatter makes the luminosity less dependent on halo mass, which makes the luminosity function flatter). On the other hand, it may be possible to fit the luminosity function with a more complicated distribution of \Lya luminosities, for example a bimodal distribution (i.e.\ a ``duty cycle'' scenario).

The simulation of the IGM can also affects the results. The simulation we used here only distinguishes between two types of sources of ionizing radiation (big and small). Changing the source model would change the IGM-induced clustering to some extent. For example, we could imagine an extreme scenario where all dark matter haloes are hosts to sources of equal ionizing flux. In this case, the IGM-induced clustering would be relatively small, since all galaxies would ionize the region around them to the same extent. On the other extreme, if there were only a few very bright ionizing sources, the clustering signal would be stronger since the observability of a \Lya emitter would depend strongly on whether it was located inside one of the few large \hii regions produced by a massive source. However, in most realistic models, each \hii region is produced by tens or hundreds of sources, and so for a fixed global ionization fraction, the results will not be strongly dependent on the relative efficiencies of different sources.

In addition to the model dependence, Fig.\ \ref{fig:count_in_cells} shows that cosmic- and sample variance add a significant degree of uncertainty. However, if the test is performed as a simple null hypothesis test (i.e.\ asking the question ``is the IGM at $z=6.5$ consistent with fully ionized IGM or not?''), model dependence is no longer a big issue. 

\subsection{HSC observations with \Lya and UV luminosities}
\label{sec:lya_plus_uv}

We now consider the much more optimistic case where we have measurements of not only the \Lya luminosity, but also the \Lya EW (which requires detections in the UV continuum as well). This is unlikely to be true for all galaxies after the initial HSC narrow-band survey, but it may be possible to perform follow-up observations to detect the UV continuum level in a larger number of galaxies. For this section, we assume---optimistically---that we know the EW for all the galaxies in the HSC sample. As such, our results in this section represent an upper limit on the statistics that will be possible using EWs.

Observations have shown no correlation between \Lya luminosity and EW \citep{zheng2013}, and we would expect that in a fully ionized IGM, the EW of an object would be independent of its position on the sky. 
However, since the EW will be attenuated by \hi absorption, a patchy ionization structure of the IGM would introduce a position dependence in the EW distribution. A possible way of looking for signs of reionization would therefore be to test whether the observed EW distribution shows any signs of spatial clustering.

\begin{figure}
	\begin{center}
		\includegraphics[width=\columnwidth]{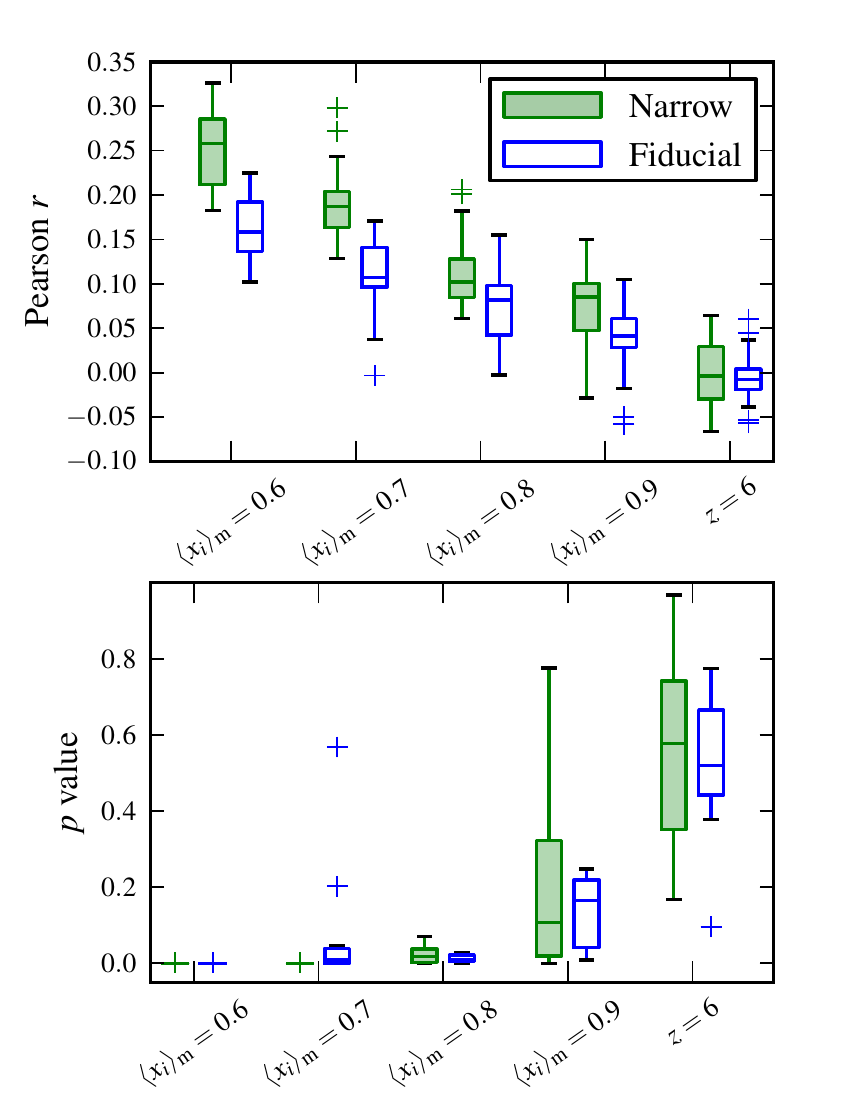}
	\end{center}
	\caption{Results from measuring the correlation between the number of neighbour galaxies and the mean EW of the neighbours (see the text for details). \emph{Upper panel}: The Pearson correlation coefficients. Notation is similar to Fig.\ \ref{fig:count_in_cells}, but here we show results for both our EW models. \emph{Lower panel}: $p$ values, indicating the likelihood of obtaining the observed correlation coefficient if the IGM is completely ionized at $z=6.5$.}
	\label{fig:rew_neigh_corr}
\end{figure}

\begin{figure}
	\begin{center}
		\includegraphics[width=\columnwidth]{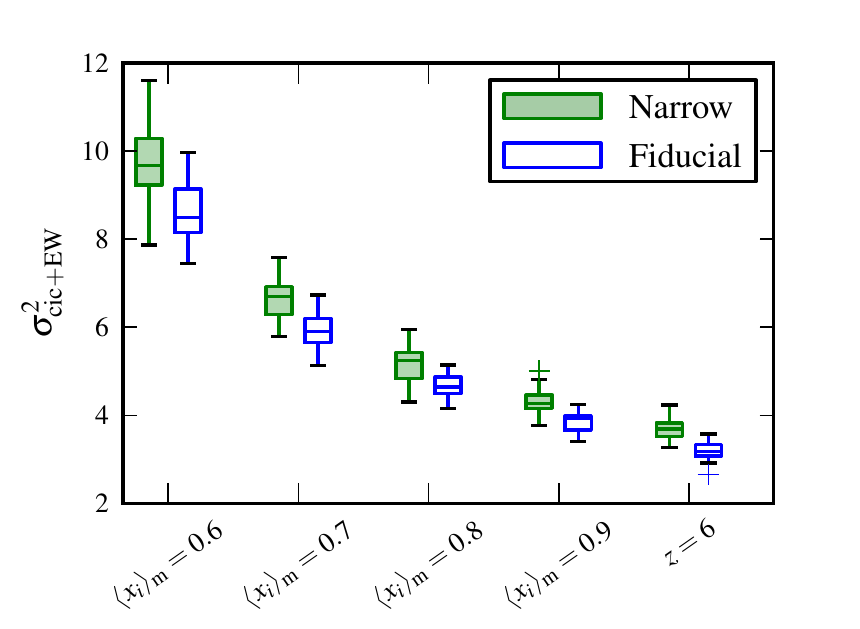}
	\end{center}
	\caption{Same as Fig.\ \ref{fig:count_in_cells}, but for counts-in-cells combined with the smoothed EW.}
	\label{fig:combo_sigma}
\end{figure}

One could think of many ways of performing such a test, but one test that we have found to perform particularly well is to measure the correlation between the number of neighbours of a galaxy and the mean EW of these neighbours. More precisely, we conduct this test as follows:
\begin{enumerate}
	\item For each galaxy: 
		\begin{itemize}
			\item Locate all galaxies within some distance $d$ on the sky-plane.
			\item Count the number of such neighbour galaxies and calculate their mean EW.
	\end{itemize}
\item Measure the Pearson correlation coefficient $r$ between the number of neighbours and the mean neighbour EW.
\item Take all measured EWs and randomly reassign them to another galaxy, to create a new sample with no correlation between EW and position on the sky. Repeat steps (i) and (ii) for the shuffled sample.
\item Repeat step (iii) many times to create a distribution of $r$ values with no correlation.

\end{enumerate}

Using this procedure we can calculate a $p$ value for the null hypothesis---that the IGM is fully ionized---by counting the fraction of shuffled samples that give a higher value of $r$ than the original sample. We have tested our method on the publicly available sample of \Lya emitters of \cite{nilsson2009}. Running our test on their $z=2$ objects (where we expect no clustering from the IGM) for a number of different values of $d$, we find no signs of correlation between the number of neighbours and the neighbour-smoothed EWs. While this sample is much smaller than our simulated volume (187 objects in a 35$\times$34 arcmin region) it is still encouraging to see that there does not seem to be any signs of intrinsic correlation in lower-redshift data.

Our results for this test on our simulated data for both of our EW models (see section. \ref{sec:simulations}), using a neighbour distance $d=20$~cMpc, are shown in Fig.\ \ref{fig:rew_neigh_corr}. For each of our scenarios we analyze the same 30 independent mock observations as in section \ref{sec:lya_only}, each with the same size as the HSC survey. We use 500 shuffled samples to calculate the distribution of correlation coefficients for the case of the null hypothesis.

As Fig.\ \ref{fig:rew_neigh_corr} shows, with a sample of this size, this method can robustly rule out a global ionized fraction lower than 80 per cent at high confidence level, at least for the narrow EW model. Already at 90 per cent ionization, it may be possible to infer some neutral IGM, albeit with a lower significance. 

\begin{figure*}
	\centering
		\includegraphics[width=17cm]{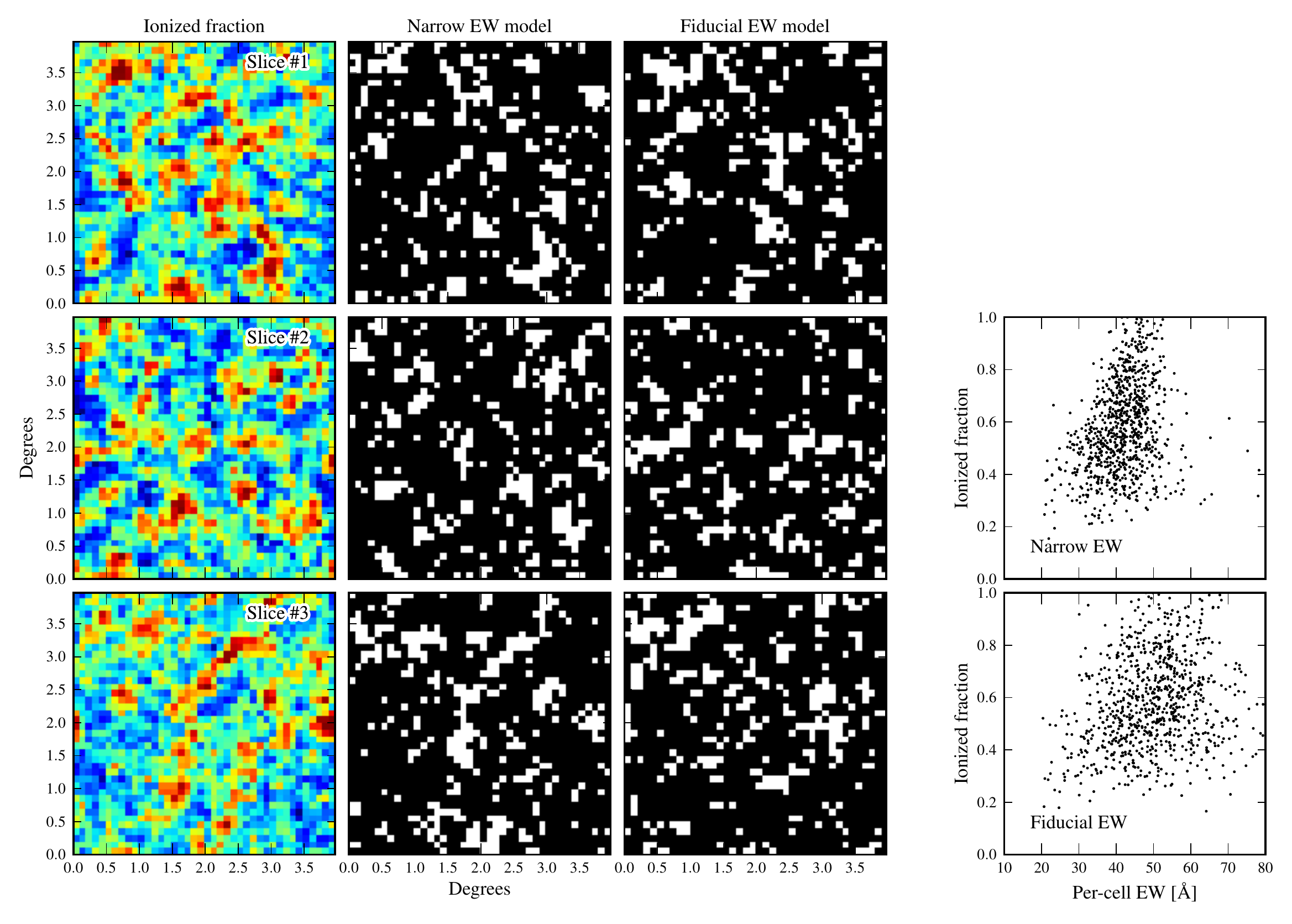}
		\caption{Bubble mapping using smoothed EWs for the $\xm=0.6$ scenario for three sample slices through the simulation volume. \emph{Left panel}: The first column shows the true ionized fraction, with blue indicating completely neutral, and red completely ionized. The second and third columns show the smoothed and gridded \Lya EWs for our two EW models. To better see the ionized regions, we show cells with EWs above the 80th percentile as white, and all other cells as black. \emph{Right panel}: Correlation between the local ionized fraction and the neighbour-smoothed EW of each cell in the third row (slice 3) of the right panel.}
	\label{fig:extracted_maps}
\end{figure*}

Besides making use of more information (since we have both UV and \Lya luminosities), this test has the advantage of being ``self-contained'', since we can directly compare to a random EW distribution. While the counts-in-cells test relies on comparing to observations at lower redshifts, the neighbours--EW test directly gives us a likelihood of the IGM being completely ionized at $z=6.5$.

It is also possible to combine the information from the EW distribution with the count-in-cells test from the previous section. To do this, we count the number of galaxies in each cell, as before, and multiply that number with the normalised average neighbour-smoothed equivalent width in that same cell. We then calculate the variance $\sigma^2_{\mathrm{cic+EW}}$ of this new quantity. The results of this test are shown in Fig.\ \ref{fig:combo_sigma}. From this figure, we see that including the EWs improves the counts-in-cells measurement slightly, to the point where a change of $\xm$ as small as 10 per cent gives a clearly detectable change in the clustering signal.

Aside from the observational difficulty involved in obtaining reliable EWs for all the narrow-band selected objects, this method has the drawback of being dependent on the intrinsic EW distribution assumed, as is clear from Fig.\ \ref{fig:rew_neigh_corr}. The wider the distribution is, the more the variation of EWs will be due to random scatter rather than varying IGM transmission. If this method was to be used to constrain the global ionized fraction for real data, one would likely want to calibrate it to better observations of the EW distribution at $z=5.7$.

\subsection{HSC bubble mapping}

A test such as the neighbours--EW correlation test described in the previous section can tell us whether there is some spatial structure to the EW distribution, but it will not provide any information about the sizes, shapes and locations of ionized regions. However, if the IGM should turn out to be significantly neutral at $z=6.5$, so that there are in fact distinct ionized bubbles in which most galaxies reside, it may be possible to extract a crude ``image'' of the ionization state of the IGM by comparing the observed EWs at different positions on the sky.

In Fig.\ \ref{fig:extracted_maps}, we have taken the objects in the $\xm=0.6$ scenario and smoothed the EWs by setting each EW to be the mean of all the EWs within $d<20$ cMpc. We have then divided the field into a grid of $40 \times 40$ cells and calculated the mean smoothed EW in each cell \footnote{The neighbour-smoothing is similar in nature to a convolution operation, and improves the results compared to only performing the gridding, since it smooths the EWs based on all the neighbours, not just those that happen to be in the same grid cell.}. We compare this to the IGM ionized fraction in a skewer extending 50 cMpc towards the observer (left column; down-sampled to the same cell size as the EWs). To better see the results we show cells with an EW below the 80th percentile as black, and everything else as white. This makes it easier to locate the regions with particularly high EWs. We show the results for three independent slices through our simulation volume for both EW models.

Although there is a large amount of noise in the ``images'', the largest ionized structures can clearly be identified for the narrow EW model (middle column). Note for example the elongated ionized structure extending from the middle to the top-right corner of slice 3, and the `J'-shaped structure in the lower-right corner of slice 1, both of which are clearly visible in the panels on the right. However, for the fiducial EW model (right column), the results are less optimistic. A few of the largest structures are still visible, but they are difficult to distinguish from the random noise.

\begin{figure*}
	\centering
		\includegraphics[width=18cm]{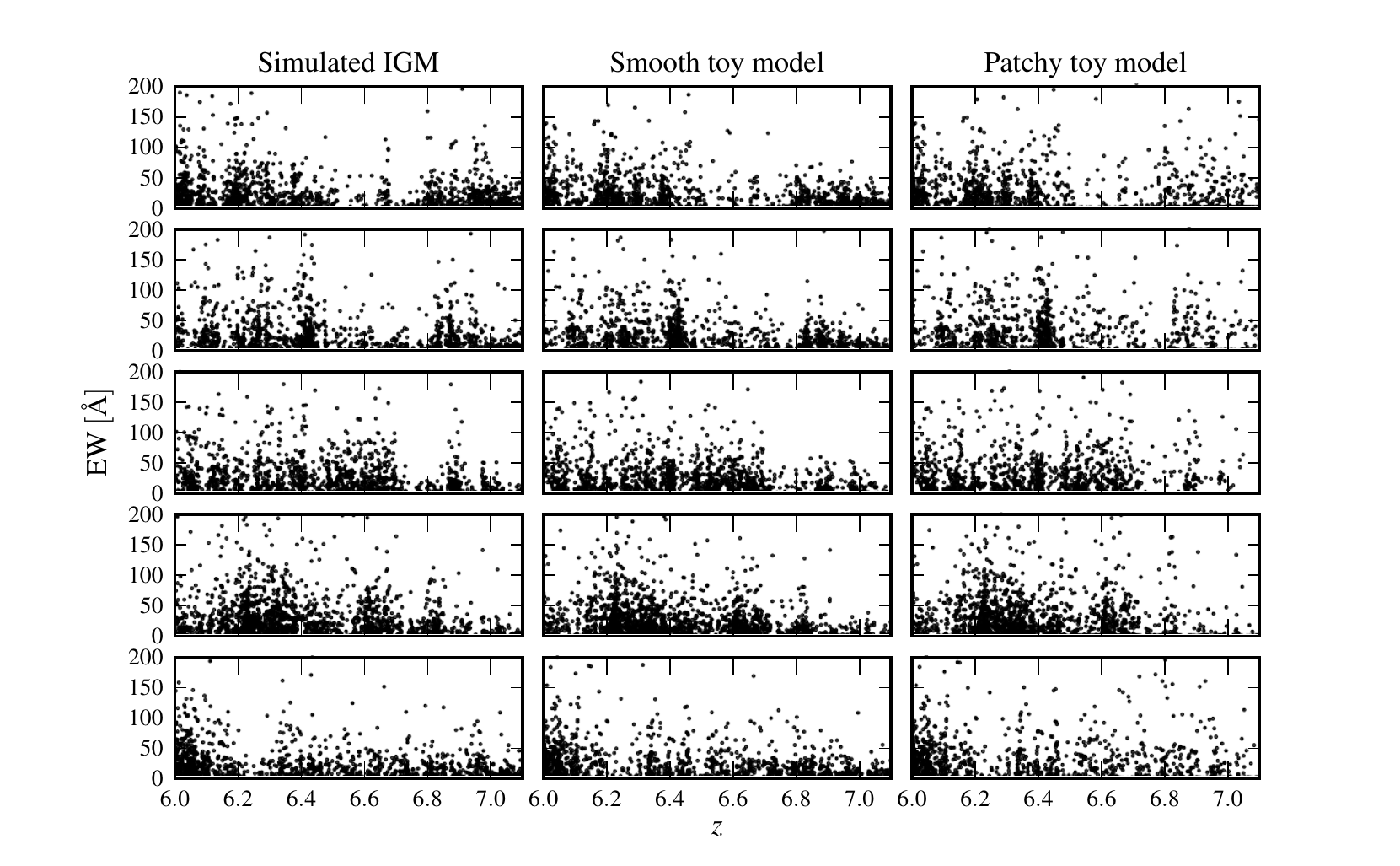}
		\caption{Redshifts and EWs of the objects in a few example skewers for the three evolution scenarios considered in this section. Each row shows one example mock observation. The values shown in these plots are the intrinsic values of the drop-out selected galaxies, not the values that would be measurable from spectroscopy.}
	\label{fig:jwst_skewers}
\end{figure*}

In the right panel of Fig.\ \ref{fig:extracted_maps} we show the mean EW in each cell in the bottom panel (slice 3) and the mean ionized fraction in front of the slice. This figure shows that while there is a large random scatter, there is a clear correlation between the EW and the local ionization state of the IGM. The correlation is significantly stronger for the narrow EW model than for the fiducial model, however, since there is more random scatter in the EWs for this model.

\subsection{JWST observations}
\label{sec:jwst}

Next, we consider a hypothetical small-field, deep spectroscopic survey, as might be carried out using the JWST with the NIRSpec instrument. We assume a survey consisting of 25 NIRSpec fields, for a total area of 225 arcmin$^2$ and a redshift range of $6<z<7$. To select drop-out targets for NIRSpec, one first needs a photometric survey reaching $m_\mathrm{AB}\approx 28.5$ mag. Reaching this limit with $S/N\approx 5$ in the F070W, F090W and F115W filters, takes $\approx 3$ h per sub-field, which implies 75~h of NIRCam imaging in total. To improve the dropout selection, auxiliary imaging at shorter optical wavelengths would also be very valuable, and HSC seems to be the ideal instrument for this---a single HSC field exposed for $\approx 14$ h would for instance be sufficient to add $m_\mathrm{AB}\lesssim 28.5$ mag data in the $g$-band (0.45 $\mu$m). Based on the luminosity functions of drop-out selected galaxies \citep{bouwens2007,bouwens2011} we estimate that there should be $\approx 1650$ targets for spectroscopic follow-up in our survey, between $6 < z < 7$. Using the NIRSpec exposure time calculator (v P1.6), and assuming an exposure time of 3 hours per sub-field (3 arcmin$\times$ 3 arcmin), we find that it should be possible to measure the \Lya EW with acceptable S/N for objects with EW$>40$ \AA, and the redshift down to an uncertainty of $\Delta z \lesssim 0.1$ for objects EW$>10$ \AA.

While the number densities that we obtain even for such an ambitious survey are much lower than those expected for the HSC survey, there are other aspects that make a survey such as this interesting to consider. Since we are measuring the spectra of objects, we can obtain their line-of-sight positions, and we are thus probing the IGM in a skewer along the redshift axis, rather than in a plane on the sky (see Fig.\ \ref{fig:lae_boxes}). Since the area on the sky of such a skewer is small compared to typical sizes of ionized regions, this type of measurement is essentially one-dimensional along the line-of-sight. This makes such measurements fundamentally different from those discussed earlier in the paper, since the survey will probe galaxies across a large range of epochs (the redshift range of our hypothetical JWST survey is $\Delta z=1$, which is 10 times that of the HSC survey). 

Previous similar surveys have focused mainly on the evolution of the fraction of drop-out selected galaxies that show \Lya emission. There is now strong observational support that this fraction drops sharply between redshifts 6 and 7 \citep{pentericci2011,stark2011,schenker2012,pentericci2014,tilvi2014}. This drop could be explained by neutral IGM, intrinsic galaxy evolution or some combination between the two. Analogous to the first part of the paper, we would like to investigate whether a larger survey may be able to break this degeneracy by looking at higher-order statistics.

To account for the change in number density of objects and IGM transmission, we construct our mock JWST observations by combining our different scenarios from section \ref{sec:scenarios}. We construct a new data volume by assuming that the IGM is 30 per cent ionized at the high-redshift end of the observations (which we put at $z=7$). This is the ionized fraction required to reproduce the results of \cite{schenker2012} and \cite{pentericci2014}, if the evolution in the EW distribution is due to neutral IGM only. We make this new data volume by taking slices of data from the different scenarios and putting together a new volume that goes from $z=6$, where the ionization fraction is 95 per cent, up to $z=7$, where the ionization is 30 per cent. For the rest of this section, we will refer to this scenario as ``Simulated IGM''. For comparison, we also construct a similar volume where the IGM is 95 per cent ionized all the way up to $z=7$. Then, we divide both of these volumes into skewers, and select the most massive objects to give the correct number density of observed drop-out galaxies (approximately 1650 objects per skewer). In total, we construct 256 different skewers, each representing a mock JWST observation.

Using the volume with a fixed, 95 per cent, ionization state we also construct two scenarios with intrinsic EW evolution. In the first of these scenarios, which we will refer to as the ``Smooth toy model'', the EWs of galaxies at a redshift $z$ are dimmed by a factor $\epsilon_s(z)$. This is equivalent to a distribution of transmitted fractions of \Lya photons that is just a delta function at a given redshift: $p(T_{\alpha}) = \delta(T_{\alpha}-\epsilon_s)$. We tune $\epsilon_s(z)$ to give the same fraction of galaxies with $\mathrm{EW} > 25$ \AA~as the scenario with neutral IGM at all $z$. 

In the second scenario, termed ``Patchy toy model'', a fraction $\epsilon_p(z)$ of galaxies at each redshift $z$ are turned off completely, while the rest are left unchanged. This is equivalent to a distribution of $T_{\alpha}$ that follows a bimodal function with one peak at $T_{\alpha}$=0 and one peak at $T_{\alpha}$=1. This model was first used by \cite{treu2012} as a simplified view of a partially ionized IGM. However, in contrast to the simulations, there is no spatial correlation in the patchiness in this model---a galaxy with $T_{\alpha}=0$ may lie right next to one with $T_{\alpha}$=1. 

One approach to separating evolution scenarios is to compare the distribution of EWs at two different redshifts. The difference in EW distribution contains more information than simply the fraction of objects that fall above some EW limit. Very recently, \cite{pentericci2014} claimed that their observations of \Lya emitters strongly favoured a patchy evolution over a smooth one between redshifts 6 and 7, and similar results were found by \cite{tilvi2014} at $z\sim8$. This was interpreted as supporting a patchy reionization as the explanation for the drop in \Lya emitter fraction.

\begin{figure}
    \begin{center}
        \includegraphics{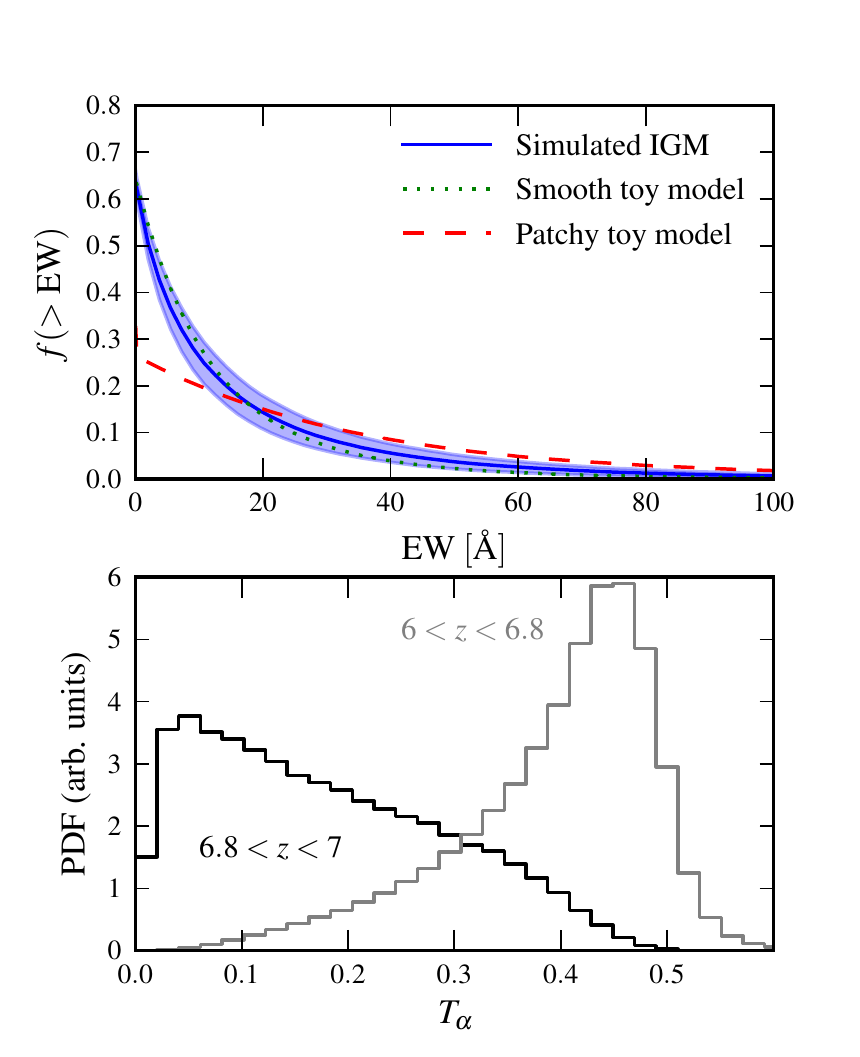}
    \end{center}
    \caption{\emph{Upper panel}: Cumulative EW distributions for all mock JWST observations for our three different evolution scenarios at $z>6.8$. The shaded blue regions shows the $1\sigma$ field-to-field variation for the Simulated IGM scenario. \emph{Lower panel}: Probability density distributions (PDFs) of the IGM transmission fraction $T_{\alpha}$ for the Simulated IGM scenario for high and low redshifts.}
    \label{fig:ew_pdf}
\end{figure}

However, as Fig.\ \ref{fig:ew_pdf} shows, our simulations (the Simulated IGM scenario) are much more similar to the smooth evolution model than the patchy model. The top panel of this figure shows the cumulative EW distribution for the high-redshift part of all our JWST-like skewers. We see that the simulations produce an EW distribution that is almost indistinguishable from the smooth evolution scenario. \cite{pentericci2014} found a few objects at $z\sim7$ with relatively high EW, but a significant shortage of intermediate-EW objects. The patchy model gives exactly such a distribution, because it is equivalent to a bimodal $T_{\alpha}$ distribution. However, as the bottom panel of Fig.\ \ref{fig:ew_pdf} shows, a neutral IGM does not produce a bimodal distribution of $T_{\alpha}$ in our simulations.

Therefore, it would seem that the results of \cite{pentericci2014} actually argue \emph{against} patchy reionization, or at least the type of patchy reionization produced by neutral IGM in our simulations. A future survey such as the one described in this section would of course provide much better measurements of the EW distribution of high-redshift \Lya emitters, but as Fig.\ \ref{fig:ew_pdf} shows, the EW distribution is not a very sensitive probe of different reionization scenarios, and is in fact not very useful for distinguishing between evolution due to neutral IGM and smooth intrinsic evolution.

As we discussed in the first part of the paper, a distinguishing feature of reionization is that $T_{\alpha}$ for a given galaxy depends to some degree on the location of that galaxy. For the mock HSC observations, we quantified this using the counts-in-cells method and the neighbours-EW correlation. Unfortunately, these types of methods are much less effective for samples that come from a wide range of redshifts. For example, in all our mock JWST observations there is a strong correlation between number of neighbour galaxies and the EW, simply because there is a higher density of galaxies at lower redshift, and the EWs of these galaxies are less attenuated, regardless of the evolution scenario.

\begin{figure}
	\centering
		\includegraphics[width=\columnwidth]{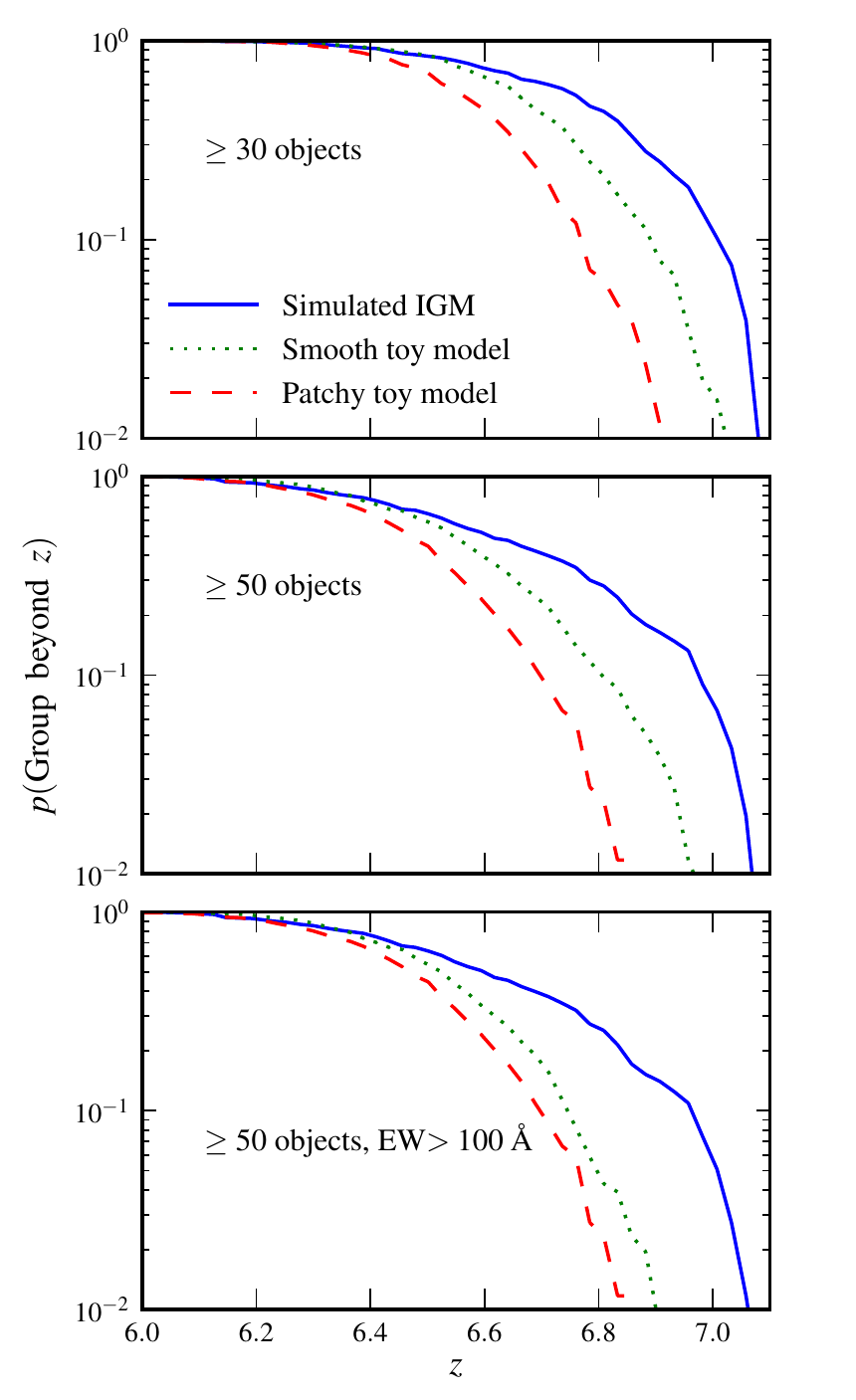}
        \caption{Probability of finding a group of more than 30 (upper panel) and 50 (middle panel) \Lya emitters beyond a given redshift for the type of JWST-like survey considered here. The lower panel shows the same as the middle panel, but now requiring that each group contain at least one object with EW$>100$ \AA.}
	\label{fig:p_galaxy_group}
\end{figure}

Looking at Fig.\ \ref{fig:jwst_skewers}, it is also clear that the Simulated IGM and the Smooth toy model scenarios are not hugely different from each other, and that the field-to-field variation is rather large. However, there does seem to be a tendency for high-redshift galaxies to be slightly more clustered in the Simulated IGM scenario, which is what one would expect if the skewer happens to cross a large ionized region. We quantify this by identifying groups of galaxies using a one-dimensional friends-of-friends algorithm along the line-of-sight. We assume that our spectroscopic survey can detect the redshifts of objects with EW$>10$\AA~with an accuracy of $\Delta z=0.1$, and identify friends-of-friends groups using a linking-length of 2~cMpc. We do this for all our 256 skewers for each evolution scenario.

In Fig.\ \ref{fig:p_galaxy_group} we show the results of this analysis. We show the probability that a single JWST observation will contain a group of galaxies of a certain size beyond a given redshift. We see here that the three evolution scenarios are fairly similar up to $z\approx6.5$. However, beyond this redshift, the Simulated IGM scenario contains significantly more groups. This difference is a direct consequence of the spatial correlation of $T_{\alpha}$. Judging by Fig.\ \ref{fig:p_galaxy_group}, finding a group of at least 50 \Lya emitters at $z>6.8$ would be a strong indication of reionization, and would be unlikely for any of the other evolution scenarios. This would only happen if the skewer happened to cross a particularly large ionized region, which is not extremely likely if the survey field is chosen at random. However, it may be possible to target fields with bright quasars, which are likely to be located in very biased regions.

We can separate the scenarios even further by requiring that the groups of galaxies contain at least one object with an EW above some threshold value. We show the results for the threshold value of $100$ \AA~in the bottom panel of Fig.\ \ref{fig:p_galaxy_group}. We see from this figure that the chances of detecting a group of at least 50 \Lya emitters beyond $z=6.8$, of which at least one has an $\mathrm{EW}>100$ \AA, is approximately 30 per cent if the evolution in EW distribution is due to neutral IGM only. This is three times as likely as the other two evolution scenarios. At $z>6.9$, there is still a 10 per cent chance of finding such a group in the Simulated IGM scenario, whereas the probability is negligible for the other two scenarios.

In summary, when it comes to using JWST observations of \Lya emitters to directly constrain the ionized fraction of the IGM, our results look somewhat less promising than for the HSC. Even a very ambitious survey like the one considered here will only be able to detect a small fraction of the objects that are detectable with the HSC. Furthermore, since the observations span a large redshift interval, any effects of the IGM will be mixed together with intrinsic evolution. Nevertheless, with a bit of luck there is a possibility that a JWST survey will detect a large group of galaxies at very high redshift. This would be a strong indication of reionization, but not a firm proof on its own, since other evolution scenarios also have a non-negligible probability of producing false positives.

Even without such luck, a survey like this will still be useful for studying reionization. What we have considered here is only the observability of the statistical effects of a partially neutral IGM on a population of \Lya emitters, but a deep spectroscopic survey will produce a great deal of information about the physical properties of the galaxies that are likely to have been the sources of reionization. Knowing more about the ages, sizes and star formation rates of these objects will be an important piece in the EoR puzzle, and deep JWST/NIRSpec observations of the type proposed here may even be able to constrain the escape fraction of ionizing radiation from some of the brightest EoR galaxies \citep{zackrisson2013}.

%%%%%%%%%%%%%%%%%%%%%%%%%%%%%%%%%%%%%%%%%%%%%%

\section{Summary and discussion}
\label{sec:summary}

Using a large-scale simulation of the IGM during reionization combined with \Lya radiative transfer simulations, we have explored the new types of statistical measurements of the ionization state of the IGM at $z\gtrsim6$ that will become possible with the next generation of large \Lya emitter surveys. Surveys with instruments such as the HSC and JWST/NIRSpec will provide improved measurements of many important properties of high-redshift \Lya emitters, such as their luminosity function, their typical sizes, ages, star formation rates etc. However, such measurements only give indirect information about the history of reionization. In this paper, we have instead focused mainly on methods to break the degeneracy between neutral IGM and intrinsic galaxy evolution. 

Clustering measurements such as counts-in-cells provide a way of separating the effects of a neutral IGM from intrinsic galaxy evolution. While changes in star formation rate, escape fraction or dust content may all help explain the apparent decrease of \Lya emitters between $z=6$ and $z=6.5$, it is unlikely that any of these processes would produce a clustering signal that could be confused with a neutral IGM. 

We find that, if only \Lya luminosities are available, the Subaru Hyper-Suprime Cam (HSC) survey should be able to distinguish an IGM that is $\gtrsim 20$ per cent neutral from one that is completely ionized at $z=6.5$ by measuring the clustering of \Lya emitters, for example using the counts-in-cells method. 

If the UV continuum luminosities can also be measured, several new possibilities open up. We have shown that by measuring the correlation between the mean EW of all the close neighbours of a galaxy and the number of such close neighbours, it is possible to detect a neutral fraction with a sensitivity similar to counts-in-cells measurements. If the two methods are combined, a neutral fraction as small as 10 per cent may be detectable at $z=6.5$ using HSC.

In this more optimistic case, it may also become possible to use the \Lya EWs to directly map the locations, sizes and shapes of the largest ionized regions. Regardless of the availability of data, such bubble mapping is of course only possible if there actually is a substantial amount of neutral hydrogen at $z=6.5$. In most reionization simulations, at an ionized fraction somewhere around 70 per cent or so, the ionized bubbles are overlapping to such a large extent that one can no longer speak of individual ionized regions. Additionally, we have seen that the feasibility of a bubble mapping measurement depends strongly on the intrinsic EW distribution---if there is a large amount of random scatter in the EWs, the correlation between local ionized fraction and observed EW is weak. If, however, such a bubble mapping measurement could be made, it could give unique insights into the topology of reionization. It would also provide great synergies with upcoming 21-cm experiments such as LOFAR and the SKA, for example for cross-correlating the \Lya emitter density with the 21-cm signal \citep{wiersma2013}.

We have also investigated the prospects for using the JWST to constrain the IGM ionization state. The statistical methods we used to look for clustering in the HSC data do not work very well for the type of data produced by a deep spectroscopic survey. First of all, the sample size will be considerably smaller than for the HSC, even for a very ambitious survey. Second, such a survey will contain objects from a wide range of redshifts, so that any effect from an increasingly neutral IGM will be mixed up with whatever intrinsic galaxy evolution effects are present.

Nevertheless, we have shown that there is a chance that such a survey will detect a group containing a large number of galaxies in a small redshift interval at high $z$. Finding such a group would be a strong indication of a neutral IGM where galaxies reside in ionized bubbles, especially if the group contains one or more high-EW objects. However, even a detection like this would not be a conclusive proof of a neutral IGM, since similar groups can occur also in the other evolution scenarios considered here, albeit with much lower probability. 

It is also interesting to compare our simulations to the recent observations by \cite{pentericci2014} and \cite{tilvi2014}, who performed spectroscopic observations of \Lya emitters at $z > 6$ and both found that the EW distribution at high redshifts is better explained by a patchy evolution model---where some objects are completely turned off and some are left as is---than a smooth model, where all galaxies are dimmed by a constant factor. The patchy model is intended as a toy model of reionization, where galaxies are either inside or outside ionized bubbles. However, in our simulations, galaxies are always located inside ionized bubbles, and the distribution of IGM transmission fractions, $T_{\alpha}$, is not bimodal. The same result is found in most similar simulations (see e.g.\ \cite{mcquinn2007} and \citealt{dijkstra2011}). With a unimodal $T_{\alpha}$ distribution, the EW distribution looks almost indistinguishable from the smooth evolution model, but very different from the patchy evolution model.

Interestingly, this implies that the results of \cite{pentericci2014} and \cite{tilvi2014} actually argue \emph{against} the reionization scenarios produced by most simulations. It is not obvious what kind of scenario could produce the observed results, which show a strong deficit of intermediate-EW objects. It is conceivable that a situation like this could be produced in a scenario like that proposed by \cite{bolton2013}, in which the IGM at high $z$ is highly ionized, but \Lya emitters are obscured by small, high-density clumps of neutral gas. If there are relatively few clumps, there could be some galaxies that are unobscured and others that are hidden behind clumps, leading to a bimodal $T_{\alpha}$ distribution. It is also possible that some change in the intrinsic properties of galaxies (such as evolution in the escape fraction; \citealt{dijkstra2014}) is strong in some group of galaxies, and weak in the rest. Observations with the JWST will provide much better statistics on the EW distribution of high-$z$ \Lya emitters. Regardless of whether the results of \cite{pentericci2014} and \cite{tilvi2014} are verified or not, this will teach us something about the evolution of galaxies in the early Universe.

While our results overall look rather promising, at least for the HSC, there are a few caveats to bear in mind. First, the observations we are considering are all very challenging, and some of our assumptions about them are somewhat optimistic. For example, obtaining UV luminosities for all the objects in the HSC survey may not be possible. Second, there are a number of uncertain parameters in our model for \Lya emitters, as discussed in more detail in section \ref{sec:lya_only}. Performing accurate simulations of \Lya production and radiative transfer through the circum- and intergalactic medium is a highly challenging task with many pitfalls. Our aim in this study has been to investigate the types of measurements that may become possible with the next generation of \Lya emitter surveys, rather than to explore every possible model uncertainty. We have therefore attempted to construct a model with reasonable assumptions that tend to lie on the conservative side.

In any case, the results from the next generation of \Lya emitter surveys will doubtless be of great value for studying reionization, regardless of what is actually observed. If a clustering signal can be seen in the HSC observations at $z=6.5$, this will be strong evidence that reionization was still ongoing at this redshift. Conversely, if no clustering is seen, this will indicate that reionization is already almost completely finished by this redshift, and instead the intrinsic properties of \Lya change rapidly between $z=6$ and $z=6.5$. For the JWST, if a large group of galaxies could be detected at high redshift, this would be a strong indication of a partially neutral IGM. However, even without such luck, a deep spectroscopic survey with the JWST would put unprecedented constraints on the evolution of the EW distribution, which contains valuable information about the first galaxies in the Universe.

%%%%%%%%%%%%%%%%%%%%%%%%%%%%%%%%%%%%%%%%%%%%%%

\section*{Acknowledgements}
G.M. is supported in part by Swedish Research Council grant 2012-4144. E.Z. acknowledges research funding from the Swedish Research Council, the Swedish National Space Board and the Wenner-Gren Foundations.
The authors thank Paul Shapiro and Kyungjin Ahn for the use of the 425$/h$ Mpc reionization simulation results. Those results were obtained in part through an allocation of advanced computing resources provided by the National Science Foundation (P.I. P. R. Shapiro) at TACC and the National Institute for Computational Sciences (NICS), and in part at the GPC
supercomputer at the SciNet HPC Consortium. SciNet is funded by: the Canada Foundation for Innovation under the auspices of Compute Canada; the Government of Ontario; Ontario Research Fund – Research Excellence; and the University of Toronto.
PL acknowledges support from the ERC-StG grant EGGS-278202.
The Dark Cosmology Centre is funded by the DNRF. M.\ H.\ acknowledges the support of the Swedish Research Council, Vetenskapsr{\aa}det and the Swedish National Space Board.
We are grateful to Masami Ouchi for useful discussions and for providing the expected number densities for the HSC survey, and to James Rhoads for valuable suggestions.

\footnotesize{
\bibliographystyle{mn2e} \bibliography{refs}
}

\label{lastpage}
\end{document}